\documentclass[
    prd, twocolumn,
    superscriptaddress, nofootinbib, amsmath, amssymb,
    aps, floatfix
]{revtex4-1}

\usepackage{graphicx,xcolor,setspace}
\usepackage[
    colorlinks=true,
    linkcolor=blue,
    urlcolor=blue,
    citecolor=blue
]{hyperref}

\usepackage[capitalise]{cleveref}
\usepackage{siunitx}
\usepackage{color,epsfig}
\usepackage{ifpdf}
\usepackage{amsmath}
\usepackage{bm}
\usepackage[english]{babel}
\usepackage{amsfonts}%
\usepackage{amssymb}
\usepackage{braket}
\usepackage{hyperref}
\usepackage{enumerate}
\usepackage{cancel}
\usepackage{multirow}
\usepackage{tikz}
\usepackage{tabularx}
\usepackage{feynmp-auto}

\usepackage{ulem}
\usepackage{array}
\usepackage{ulem,fancyvrb}
\usepackage{xcolor}

\definecolor{nicered}{rgb}{0.7,0.1,0.1}
\definecolor{nicegreen}{rgb}{0.1,0.5,0.1}
\hypersetup{colorlinks,citecolor= nicegreen,linkcolor= nicered}

\usepackage{dcolumn}
\usepackage{bm}
\usepackage{slashed}

\def\comments{true}

\ifx\comments\undefined
	\newcommand{\comment}[1]{}
\else
	\newcommand{\comment}[1]{#1}
\fi

\definecolor{maroon}{cmyk}{0,0.87,0.68,0.32}

\begin{document}

\title{MeV Gamma-Ray Constraints for Light Dark Matter from Semi-Annihilation}

\author{Jun Guo}
\email{jguo\_dm@jxnu.edu.cn}
\affiliation{College of Physics and Communication Electronics,
Jiangxi Normal University, Nanchang 330022, China}

\author{Lei Wu}
\email{leiwu@njnu.edu.cn}
\affiliation{Department of Physics and Institute of Theoretical Physics, Nanjing Normal University, Nanjing, 210023, China}

\author{Bin Zhu}
\email{zhubin@mail.nankai.edu.cn}
\affiliation{Department of Physics, Yantai University, Yantai 264005, China}

\begin{abstract}
Exploring the realm of Dark Matter research, Light DM, which has a mass in the range of 1 MeV to 1 GeV, is a fascinating topic both theoretically and experimentally.  We assume that the light dark matter is composed of complex scalars and produced from semi-annihilation, which is close to the scale of the MeV Gamma-ray satellite, allowing us to explore the implications of this hypothesis. The experimental data we used to constrain the scenario is from five different sources: the  COMPTEL, EGRET, INTEGRAL, Fermi Gamma-ray Space Telescope, and the e-ASTROGAM future reach. We use the analytical formula to measure the X-ray spectra allowing us to determine the annihilation cross-section bounds from $10^{-28}\mathrm{cm}^3/\mathrm{s}$ to $10^{-22}\mathrm{cm}^3/\mathrm{s}$ for different combinations of dark matter and mediator masses. We found that the MeV gamma-ray provides valuable insight into the structure of the semi-annihilation dark matter, where EGRET contributes to the stringent constraint to the semi-annihilation, and the e-ASTROGAM future reach could probe the whole parameter space of the model. 
\end{abstract}

\maketitle

\section{Introduction}
Dark matter is a hypothetical form of matter that cannot be seen but must exist because of the gravitational effects it has on our universe and the formation of galaxies. It is widely believed that dark matter is composed of particles that do not emit light or other radiation, which accounts for about $27\%$ of the total mass and energy in the universe and is difficult to detect by conventional techniques. Due to the fact that we are unsure of their composition and behavior, they rank among the most significant physics riddles of the present.

There are several possible candidates for dark matter, but the most theoretically appealing ones are the WIMPs (Weakly Interacting Massive Particles)~\cite{Bertone:2004pz} since their production mechanisms are naturally related to thermal equilibrium. The relic density of dark matter is determined by the freeze-out process occurring when dark matter particles become so rare that their number no longer changes significantly with time evolution. By computing the Boltzmann equation describing the evolution of dark matter number density, we can determine the amount of dark matter present in the universe today. These particles are massive and interact only weakly with the visible sector, making them very difficult to detect. There are some approaches to searching for them, including direct detection~\cite{Gaitskell:2004gd} and indirect detection~\cite{Slatyer:2017sev}, both of which provide a robust constraint to WIMPs~\cite{Arcadi:2017kky}. Direct detection experiments are proposed to search for the scattering of WIMPs off nuclei in detectors, and Indirect detection experiments search for the products of WIMP annihilations, such as gamma rays, positrons, and anti-protons. The parameter space that typically governs dark matter self-annihilation also dictates the dark matter-nucleon scattering cross-section. However, the dark matter-nucleon scattering cross section is now severely restricted by direct detection experiments, which rules out most of the available parameter space of WIMP annihilation~\cite{Roszkowski:2017nbc}. 

One easy way to weaken the direct detection bounds is to decrease the dark matter mass~\cite{Lin:2022hnt,Kahn:2021ttr}, as the Direct detection constraint is not applicable when the recoil energy is smaller than the detector threshold. It is because the conventional detection approach relies on detecting the tiny amounts of energy deposited by DM via nuclear recoils, which is rendered useless for DM considerably lighter than a typical nucleus. Therefore, efforts are being made to create novel detection techniques, such as the application of new targets~\cite{Essig:2011nj,Hochberg:2015pha,Essig:2015cda,Hochberg:2015fth,Schutz:2016tid,Derenzo:2016fse,Budnik:2017sbu,Cavoto:2017otc,Trickle:2019ovy,Blanco:2019lrf,Prabhu:2022dtm} or novel processes~\cite{Kouvaris:2016afs,Ibe:2017yqa,Bringmann:2018cvk}, most of which belong to direct detection, while only a few pieces of research~\cite{Essig:2013goa,Bhattacharjee:2022lts,Caputo:2022dkz,Coogan:2021sjs,Cirelli:2020bpc} focus on the indirect detection of light dark matter.

Another strategy is to disentangle direct detection observables from the dark matter relic density by dissolving the link between dark matter annihilation and scattering cross-section. The link is based on the widely held belief that Dark Matter is stabilized by a $Z_2$ parity. However, this situation is not generic and should not be a rule of thumb. Each scenario other than $Z_2$ parity should be evaluated on its own merits, among which the $Z_3$ parity is a natural and minimal extension. Processes that involve an odd number of dark matter fields can be found without leading to DM decay. We have only one sort of process if we restrict ourselves to $2\rightarrow 2$ annihilations, semi-annihilation~\cite{DEramo:2010keq}, which has been probed in the context of indirect detection signatures for GeV-scale DM particles~\cite{DEramo:2012fou,Queiroz:2019acr}. To our best knowledge, there is no research on the light dark matter produced by semi-annihilation in the literature.

The main idea of this paper is to fill the gap, where we use the MeV gamma-ray as a probe for the properties of light dark matter in semi-annihilation. The two main and novel ingredients are: (i) the dark matter belongs to sub-GeV dark matter, which has never been considered in semi-annihilation; Over the years, most of the experiments conducted in particle physics have focused on either Weakly Interacting Massive Particles (WIMPs) or axions, leaving out other potential alternatives which are equally valid and justified. While these two particles have been the primary focus of research, it is crucial to recognize the value of exploring other avenues of injury. (ii) the semi-annihilation mechanism lacks a definite signature in the MeV gamma-ray search in the literature. The answer is that the sensitivity of these experiments such as Fermi-LAT loses sensitivity. Indeed, research has already demonstrated the efficacy of utilizing the data to constrain Light Dark Matter (LDM) models. As a result, scientists have been able to narrow down the range of possible LDM, and thus gain a better understanding of the nature of dark matter, which has been a step in the effort to learn more about the Universe.
 
We will show that such an approach generates the most severe constraints to the semi-annihilated light dark matter. We mention that we do not consider the semi-annihilation process involving SM particles in the final state, since dark matter is sub-GeV, thus forbidding the possibility of the Z and Higgs boson final state kinematically. The canonical realization is to include a scalar or vector mediator, that could couple the standard model particles and dark matter simultaneously. In this paper, we choose a scalar Higgs portal mediator and scalar dark matter as a representative framework, while other possibilities of different spin of dark matter and mediator are easy to generalize.

\section{Semi-annihilation dark matter: Models, Relic Density and Direct Detection Constraints}
\label{SecClock}
\subsection{$Z_3$ dark matter model}
Our dark matter (DM) model is motivated by $Z_3$ symmetry and its semi-annihilation mechanism. Even though there are lots of previous introductions~\cite {Belanger:2012vp, Belanger:2014bga, DEramo:2010keq, Guo:2021rre}, our work is focusing on MeV scale DM, so we 
first give a brief introduction to our $Z_3$ DM model. The minimal form of the well-known $Z_3$ DM model contains only one complex scalar $S$, but in this work, we focus on the MeV scale
DM, to make DM annihilate efficiently, we introduce another scalar field $\Phi$, which is $Z_3$ singlet and gauge singlet. Under these assumptions, our model has the Lagrangian containing the following 
terms:
\begin{align}
 -{\cal L}_{Z_3}\supset &M_{s}^2S S^* + \lambda_{sh}|S|^2|H|^2+ \lambda_{s\phi}|S|^2|\Phi|^2 
\\ \nonumber
&+\left( \frac{A_s S^3}{3} + \lambda_{h\phi} |H|^2 |\Phi|^2 + c.c\right)
\end{align}
The DM candidate receive mass term from bar mass term $M_{S}^2S S^*$, Higgs portal term 
$\lambda_{sh}|S|^2|H|^2$ and $\Phi$ coupling term $\lambda_{s\phi}|S|^2|\Phi|^2$. Since the
strong limitation of Higgs invisible decay, we suppress the Higgs-portal coupling term by hand, so the squared mass of DM is
\begin{align}
m_{S}^2 \simeq M_s^2 + \lambda_{s\phi}v_{\phi}^2,
\end{align}
where we have written $\Phi = \phi + v_{\phi}$. 

The $Z_3$ singlet $\phi$ couples with SM Higgs through gauge invariant term $\lambda_{h\phi} |H|^2 |\Phi|^2$, resulting $\phi$ mixing
with the Higgs. After diagonalizing the $\Phi$/H mixing matrix, we can replace $\phi$ and h
with eigenstates in the form:
\begin{align}
h \rightarrow h\cos\theta  - \phi\sin\theta \ \ \ \ \ \ \ \ s \rightarrow h\sin\theta  + \phi\cos\theta
\end{align}
with mixing angle $\theta$. This results in couplings between $\phi$ and SM particles in the form:
\begin{align}
\label{eq:phiterms}
-{\cal L}_{\phi}\supset &+\sin\theta \sum_f\frac{y_f}{\sqrt{2}}\phi\bar{f}f + 3\sin\theta\frac{\alpha_{EM}}{4\pi}\frac{\phi}{\Lambda} F_{\mu\nu}F^{\mu\nu}  
\\ \nonumber
&-\frac{5}{6}\sin\theta \frac{\alpha_{s}}{4\pi}\frac{\phi}{\Lambda}G^a_{\mu\nu}G^{a\mu\nu}
\end{align}
the last two terms need to integrate out SM particles~\cite{Marciano:2011gm},
where $\Lambda$ is the cut-off scale of the theory, which usually equals $v_h=$ 246 GeV. The interaction terms in Eq.~\ref{eq:phiterms} help the decay of $\phi$, since our model focus on MeV-scale DM phenomenology, the mass of $\phi$ shall be $\sim {\cal O}(100)$ MeV, which means $\phi$ could only decay into some light particles (such as photon, electron, muon, and pions) and $m_{\phi}$ will give some direct affection on DM indirect detection.

The semi-annihilation channel we concentrate on is $S S \rightarrow S^* \phi$ contributes by the Lagrangian term $\lambda_{s\phi}|S|^2|\Phi|^2 $,

\begin{center}
\begin{figure}[htbp]
\centering
\includegraphics[width=0.25\textwidth]{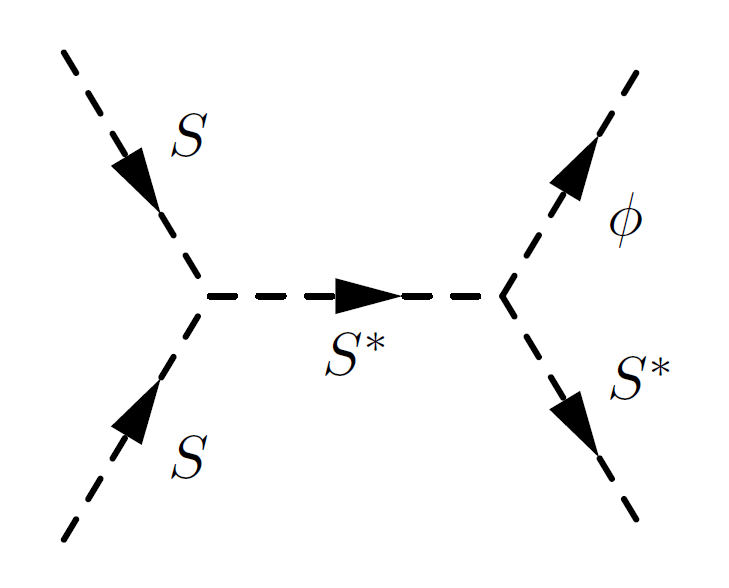}
\label{fig:m_phi2}
\end{figure}

\end{center}
to open such a channel, the mass relationship $m_{\phi} \leq m_{S}$ should
be satisfied, and the cross-section is proportional to $\propto A_s^2\lambda_{s\phi}^2v_{\phi}^2$. But at the same time, $S S^*\rightarrow \phi\phi$ channel with cross-section proportional to $\propto \lambda_{s\phi}^2$ will open since we only concentrate on semi-annihilation in this work, it is necessary to turn off Higgs-portal coupling and suppressing $\lambda_{s\phi}$. At the same time, a large enough $A_S$ will allow us to get the correct relic density without opening the Higgs-portal sector and double $\phi$ final state annihilation channel. 
Compare with the usual Higgs/$\phi$-portal DM model~\cite{Krnjaic:2015mbs} in which annihilation processes include double Higgs or $\phi$, the semi-annihilation feature of our model will leave some impact on both DM relic density and  indirect detection gamma-ray experiment because semi-annihilation will only contribute half contribution to 
the DM effective annihilation cross-section, and for the different masses between final states, the boosting level is different from the usual case, which results in a different $\phi$ decay photon spectrum.

The free parameters of our semi-annihilation DM model are Higgs mixing angle $\sin\theta$, 
mass of DM $m_S$, mass of mediator $m_{\phi}$, $Z_3$ term coupling $A_S$ and 
coupling of DM-mediator $\lambda_{s\phi}$. For simplicity, we treats $A_S v_{\phi}$ as one parameter $g_{s\phi}$, since the semi-annihilation cross section is directly
proportional to $\propto \lambda_{s\phi}^2g_{s\phi}^2$, and we need analysis $\lambda_{s\phi}$ alone to suppress $S S^*\rightarrow \phi\phi$ cross-section.

\subsection{Relic Density}
\begin{figure}[htbp]
\centering
\includegraphics[width=0.4\textwidth]{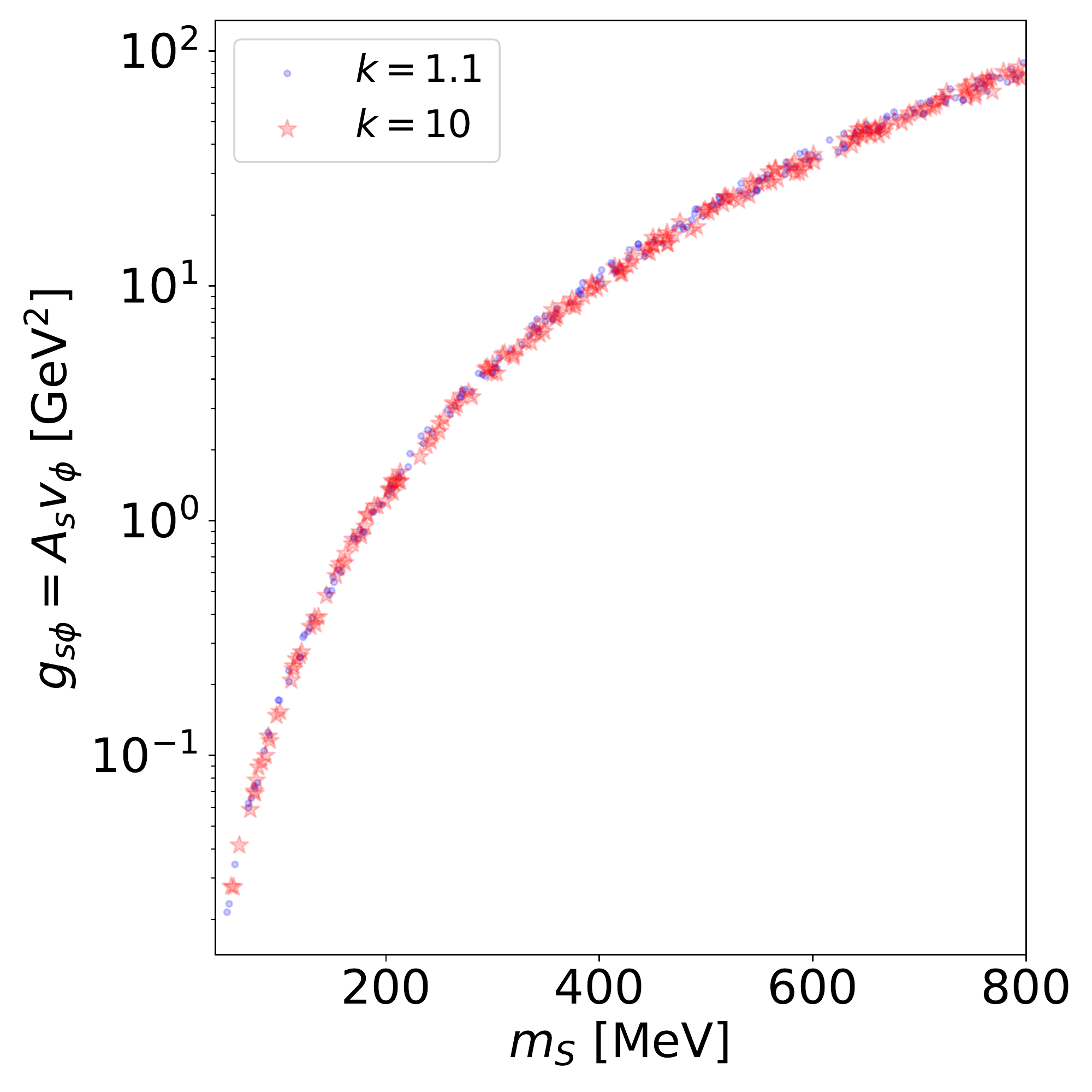}
\caption{All points satisfy the strict relic density constraint $0.094\leq\Omega h^2\leq 0.129$. We can see
the value of $g_{s\phi}=A_sv_\phi$ is in our estimated range.}
\label{fig:asvphi}
\end{figure}

Take $Z_3$-symmetric theory as a thermal freeze-out target, which suggests a single candidate for dark matter $S$ as usual. The novel aspect is the new semi-annihilation process following the Boltzmann equation to describe the evolution of dark matter number density,
\begin{align}
\frac{dY}{dt}=-s \langle\sigma v\rangle\left(Y^2-r Y \bar{Y}-(1-r) \bar{Y}^2\right)
\end{align}
where the yield $Y=n/s$ is the ratio between number density and entropy $s$, $\langle\sigma v\rangle$ is the combination of the thermally averaged cross-section for direct annihilation $SS^{*}\rightarrow\phi\phi$ and semi-annihilation $S S\rightarrow S^{*}\phi$ process,
\begin{align}
\langle\sigma v\rangle \equiv\left\langle \sigma^{S S^* \rightarrow \phi \phi}v\right\rangle+\frac{1}{2}\left\langle  \sigma^{S S \rightarrow S^* \phi}v\right\rangle
\end{align}
with the fraction $r$ being, 
\begin{align}
    r=\frac{1/2 \left\langle  \sigma^{S S \rightarrow S^* \phi}v\right\rangle}{\langle\sigma v\rangle}
\end{align}
Here $r=1$ corresponds to the pure semi-annihilation process. 
The semi-annihilation cross-section times DM relative velocity of our case is given as:
\begin{align}
\sigma^{S S \rightarrow S^* \phi}v &= \frac{1}{64\pi}\frac{|\vec{p}_{\phi}|A_s^2(\lambda_{s\phi}v_{\phi})^2}{9m_S^7}
\end{align}
where $|\vec{p}_{\phi}|$ is the momentum of final state $\phi$ in the center-of-mass frame, we may represent it
in the form:
\begin{align}
|\vec{p}_{\phi}| \simeq m_S\lambda(1, m_S^2/4m_S^2, m_{\phi}^2/4m_S^2) \label{eq:1phiS}
\end{align}
with $\lambda(1, x, y) = \sqrt{(1-x-y)^2-4xy}$, which usually results in a moderate phase space
suppression. At same time, the interaction term $\lambda_{s\phi}|S|^2|\Phi|^2$
will contribute annihilation process $S S^* \rightarrow \phi \phi$ with:
\begin{align}\label{eq:phiphi}
\sigma^{S S^* \rightarrow \phi \phi}v&=\frac{\lambda_{s\phi}^2}{32\pi m_S^2}\frac{|\vec{p}_{\phi}|}{m_S} \simeq\frac{\lambda_{s\phi}^2}{32\pi m_S^2}\sqrt{1-m_{\phi}^2/m_S^2}\\ \nonumber
&\simeq \frac{\lambda_{s\phi}^2}{32\pi}\frac{1}{0.3^2}\left( \frac{0.3}{m_S}\right)^2\simeq 0.1\times\lambda_{s\phi}^2\ \  \rm GeV^{-2}
\end{align}
Since we only concentrate on semi-annihilation with $r\simeq 1$ in this work, the main contributed channel of getting correct relic density is $S S\rightarrow S^* \phi$, which means:
\begin{align}\label{eq:estimate}
\sigma^{S S^* \rightarrow \phi \phi} \ll 1\times 10^{-8}\ \  \rm GeV^{-2}\\ \nonumber
\sigma^{S S \rightarrow S^* \phi} \simeq 1\times 10^{-8}\ \  \rm GeV^{-2}
\end{align}
from Eq.\ref{eq:phiphi}, we may set $\lambda_{s\phi}\simeq 10^{-5}$, and according to Eq.\ref{eq:1phiS}, we can estimate semi-annihilation thermal average cross section as:
\begin{align}
\sigma^{S S \rightarrow S^* \phi}v &\simeq \frac{1}{64\pi}\frac{2\times 10^{-8}}{9\times 0.3^6}\left(\frac{0.3}{m_S}\right)^6\left(\frac{A_s\lambda_{s\phi}v_{\phi}}{2\times 10^{-4}}\right)^2 \\ \nonumber
&\simeq 1.5\times 10^{-8} \left(\frac{0.3}{m_S}\right)^6\left(\frac{\lambda_{s\phi}g_{s\phi}}{2\times 10^{-4}}\right)^2\ \ \rm GeV^{-2}
\end{align}
which means $\lambda_{s\phi}g_{s\phi}\simeq 10^{-4}$, resulting $A_S v_{\phi}$ of order ${\cal O}(10) \ \rm GeV^2$. We display the numerical result by using micrOMEGAs~\cite{Belanger:2020gnr}. To learn how well 
the model fits the relic density requirement $0.094\leq\Omega h^2\leq 0.129$~\cite{2013ApJS20819H}, we scan the parameter space with
Metropolis-Hastings algorithm~\cite{mickay}, we study two extreme cases, the mass degenerate case with 
$k=m_S/m_\phi=1.1$, and light mediator case with $k=m_S/m_\phi=10$. Despite the fact $\theta$ will not affect DM relic density, but according to ~\cite{Balaji:2022noj}, SN1987A excludes $1.0\times10^{-7} \lesssim \sin\theta \lesssim 3.0\times 10^{-5}$ and scalar mass up to 219 MeV. We set the mixing angle in the range $\sin\theta \in [5\times 10^{-5}, 10^{-4}]$, for there still exists a sizeable non-excluded area according to the right panel of Fig.5 in ~\cite{Balaji:2022noj}. In Table.~\ref{table:param}, we show the range and step size of the parameters. The numerical result is shown in Fig.\ref{fig:asvphi}.

\begin{center}

 \begin{tabular}{|c|c|c|} 
  \hline
  parameter & range & step size \\
  \hline
   $\lambda_{s\phi}$ & $1\times 10^{-5}$ & $0$\\
  \hline
   $\log_{10}g_{s\phi}$ & $[-2, 2]$ & $0.3$\\
  \hline
   $m_{S}$ & $[50, 800]\rm\ MeV$ & $300 \rm\ MeV$ \\
  \hline
   $k = m_{S}/m_{\phi}$ & 1.1, 10 & $0$\\
  \hline
  $\sin\theta$ & $[5\times 10^{-5}, 10^{-4}]$&  $2\times 10^{-5}$\\
  \hline
 \end{tabular}  
 \end{center}
 \label{table:param}

\section{Precision Calculation on MeV Gamma-ray}
Gamma-ray photon intensity is given in the following form:
\begin{align}
\frac{d\Phi}{dE} = \frac{\langle\sigma v\rangle}{8\pi f_{\rm DM}m_S^2}\frac{dN}{dE} J
\end{align}
it shows the gamma-ray flux generated by DM annihilation inside the Galactic halo depends on the thermal average 
annihilation cross section $\langle\sigma v\rangle$, DM mass $m_S$, gamma-ray spectrum
$\frac{dN}{dE}$ and J-factor, where $f_{\rm DM}$ is 2 for $S$ is not self-conjugate.

\begin{figure}[htbp]
\centering
\includegraphics[width=0.4\textwidth]{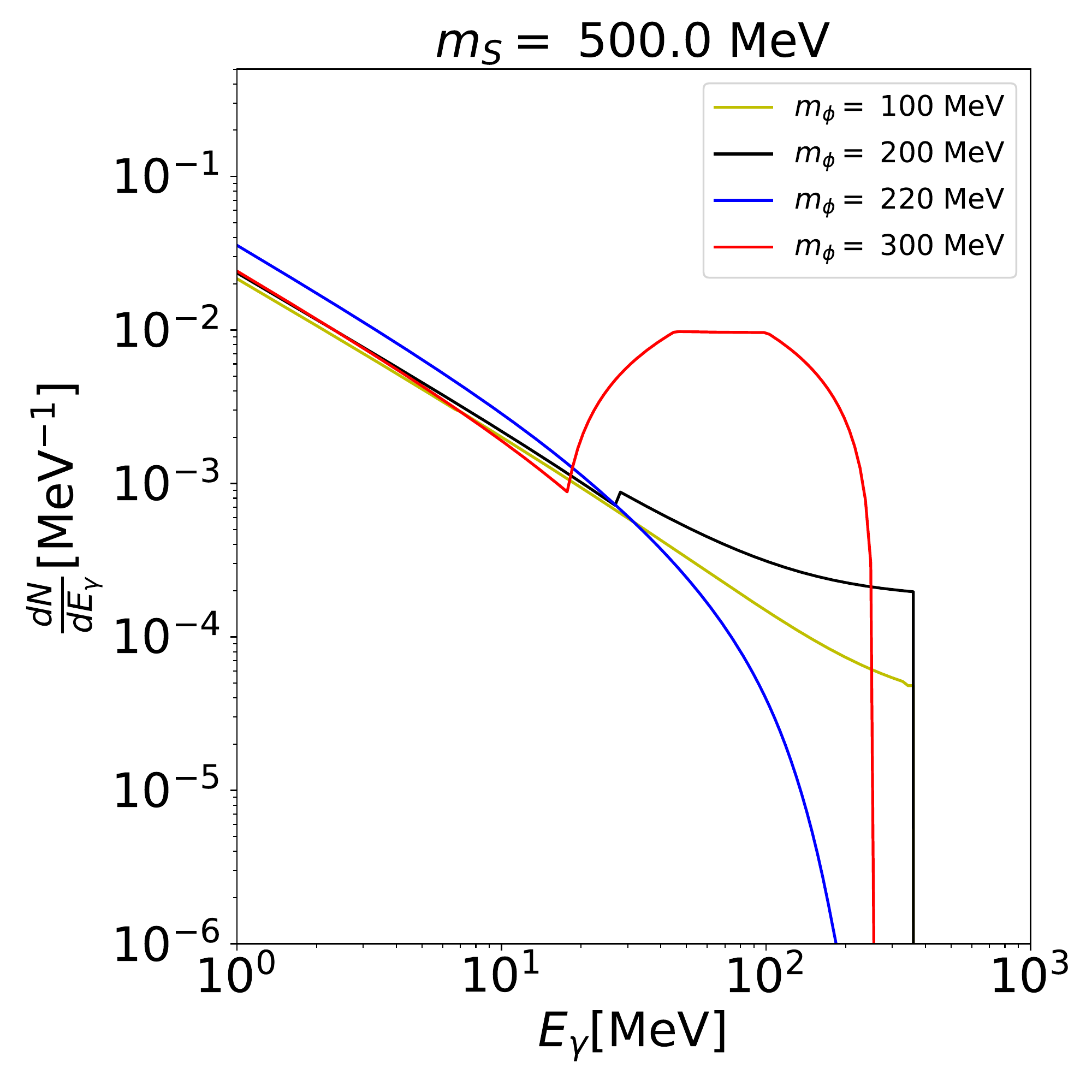}
\caption{Gamma-ray spectrum from DM semi-annihilation, generated by Hazma. In all cases with DM mass
$m_S = 500$ MeV, We find spectrum generate by case $m_\phi=200$ MeV $ < 2m_\mu$ generate less high energy photon signal. The spectrum is highly $m_\phi$ dependent.}
\label{fig:m_phi1}
\end{figure}

The Gamma-ray spectrum per annihilation depends on DM mass, mediator mass $m_{\phi}$, and coupling of $\phi$-SM. All the signal photon flux comes from the decay of the mediator
which is generated by the semi-annihilation. According to the interaction terms in Eq.~\ref{eq:phiterms}, all the $\phi$-SM couplings are proportional to $\sin\theta$, this means the 
Higgs mixing will not affect final spectrum, for the decay branching fractions will not change
with $\sin\theta$. The mass of mediator $m_{\phi}$ will determine the available decay channels of $\phi$. For a light enough $\phi$ with $m_{\phi} < 2m_{\mu}$ it can only decay into electrons with final state radiations. When $2m_{\mu}<m_{\phi} < 2m_{\pi}$, for the coupling between $\phi$ and $\mu$ is proportional to $m_{\mu}$, the 
$\mu^+\mu^-$ final state will dominant decay product of $\phi$, result in suppressing sharp photon from $e^+e^-$ final
state radiation. When $m_\phi > 2m_{\pi^{0(\pm)}}$, the decay spectrum of pions will dominate high energy spectrum region, which includes a so-called box spectrum, as shown in Fig.~\ref{fig:m_phi1}

In the usual Higgs/$\phi$-portal case, DM  will annihilate into a pair of mediator $\phi$, but in our model
only one $\phi$ is included, which means semi-annihilation will generate less photon signal, resulting in a weaker limitation from indirect detection. At the same
time, in the usual case model, the energy of final state $\phi$ is $E_{\phi}\simeq m_{S}$, but in our model, for the mass splitting between DM and mediator, the energy of $\phi$ is
\begin{align}
E_{\phi}&= \frac{E_{cm}^2-m_S^2+m_{\phi}^2}{2E_{cm}}\\ \nonumber
        &\simeq m_S -\frac{m_S^2-m_{\phi}^2}{4m_S}
\end{align}
This means, the larger the masses split, the less boosted $\phi$ is, resulting in a less energetic 
gamma-ray spectrum. In Fig.~\ref{fig:m_phi2}, we can see the usual case Higgs/$\phi$-portal
model generates almost twice the number of photon signal of our $Z_3$ model produces, and
for the different boosting result of $\phi$, compare with single $\phi$ spectrum of usual case Higgs/$\phi$-portal model, our model generate more photon 
below 1 MeV, which means less photon signal will concentrate at high energy region.
\begin{figure}[htbp]
\centering
\includegraphics[width=0.4\textwidth]{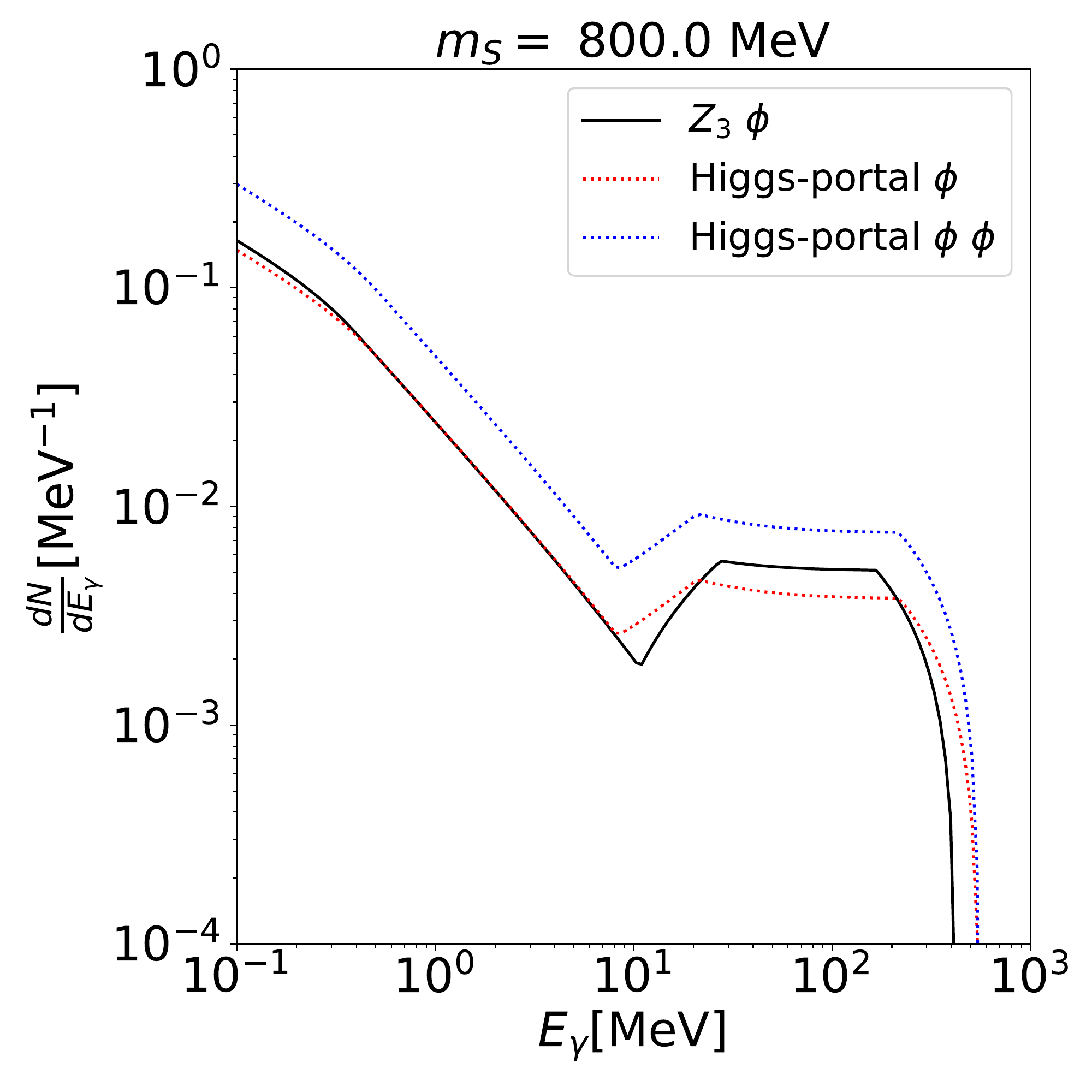}
\caption{Comparison of spectrum from our model, from usual case $\phi$-portal $\phi \phi$ final state and from one of the generated $\phi$ of $\phi$-portal $\phi \phi$ final state. All models are set with parameters $m_S=$ 800 MeV, and $m_{\phi}=$ 300 MeV.}
\label{fig:m_phi2}
\end{figure}

The J-factor contains information about dark matter density distribution in the Galactic halo, which is integrated over the observed line of sight s and solid angle $\Omega$:
\begin{align}
J = \iint_{r.o.i} d\Omega  ds \rho(r(s, l, b))^2
\end{align}
where DM density $\rho$ is a function of radial distance from Galactic center $r$, while
$r$ is given as a function of line of sight $s$, Galactic coordinate $(l, b)$ and distance of sun to Galactic center $R = 8.5$ kpc:
\begin{align}
r = \sqrt{s^2 + R^2 -2sR\cos l\cos b}
\end{align}
For the DM density distribution $\rho$, we consider the Navarro-Frenk-White(NFW) profile\cite{Navarro:1996gj}
\begin{align}
\rho_{\rm NFW}(r) = \frac{\rho_s}{(r/r_s)(r/r_s + 1)^2} 
\end{align}
and Isothermal profile\cite{Begeman,Bahcall:1980fb}.
\begin{align}
\rho_{\rm Iso}(r) = \frac{\rho_s}{1+(r/r_s)^2}
\end{align}
Following ~\cite{Cirelli:2010xx}, we set the scale factor $r_s=24.42$ kpc and $\rho_s = 0.184 \ \rm GeV\ cm^{-3}$
in the NFW profile. In the Isothermal profile, we set $r_s=4.38$ kpc and $\rho_s = 1.387 \ \rm GeV\ cm^{-3}$.

\begin{center}

 \begin{tabular}{|c|c|c|} 
  \hline
  Detector & $\Delta \Omega\ (\rm sr)$ & $\bar J_{\rm NFW(Iso)}\ (\rm MeV^2\rm cm^{-5} \rm sr^{-1})$ \\ 
  \hline
   EGRET & 6.585 & $3.78(3.64)\times 10^{27}$ \\
  \hline
   Fermi-LAT & 10.817 & $4.53(3.99)\times 10^{27}$ \\
  \hline
   Integral & 0.542 & $6.67(2.58)\times 10^{28}$ \\
  \hline
   e-ASTROGAM & 0.121 & $1.79(0.36)\times 10^{29}$ \\
  \hline
 \end{tabular}  
 \end{center}
 \label{table:J-factor}

We use \textbf{Hazma}\cite{Coogan:2019qpu} to set limits on the DM annihilation cross-section. In \textbf{Hazma}, two kinds of experiments detecting gamma-ray are implemented: the existing 
one and the upcoming one. For the existing
experiment, we choose EGRET~\cite{Thompson:1993zz} and Fermi-LAT~\cite{2009ApJ697}, while for the upcoming experiment, we choose e-ASTROGRAM~\cite{e-ASTROGAM:2016bph}. EGRET mainly focuses on gamma rays in the energy range 27 MeV - 8.6 GeV, and \textbf{Hazma} chooses the r.o.i as
$20^{\circ}<|b|<60^{\circ}$ and $|l|<180^{\circ}$. For Fermi-LAT, it focuses on gamma-ray in the energy range 150 MeV - 95 GeV, and \textbf{Hazma} chooses the r.o.i as $8^{\circ}<|b|<90^{\circ}$ and $|l|<180^{\circ}$. For the upcoming e-ASTROGRAM, the detecting energy range is 0.3 MeV - 3 GeV, which is much more sensitive than EGRET and Fermi-LAT, the 
r.o.i choosen by \textbf{Hazma} is $|b|<10^{\circ}$ and $|l|<10^{\circ}$. The averaged J-factor in r.o.i $\bar J$ values are show in table \ref{table:J-factor}, both NFW and Isothermal case are given.
In the limit setting process, 
\textbf{Hazma} use binned method for the existing experiments, it requires the flux generated by the model at any single bin not to exceed the observed value plus twice the error bar. And for the upcoming experiment, \textbf{Hazma} uses an unbinned procedure, it requires a background model, \textbf{Hazma} implements a power law background. The number of total photons generate by the DM (background) model from $E_{min}$ to $E_{max}$ obey Poisson distribution with average ratio value:
\begin{align}
\mu \simeq T_{obs}\int_{E_{min}}^{E_{max}}dE A_{\rm eff}(E) \frac{d\Phi}{dE}
\end{align}
$A_{\rm eff}$ is the detector effective area, and $\frac{d\Phi}{dE}$ is the photon spectrum generated by DM annihilation or background model. The unbinned method requires the signal-to-noise ratio to be 
significant at $5\sigma$ level, meaning $N_{\rm DM}/\sqrt{N_{\rm BG}}<5$.
\begin{figure}[htbp]
\centering
\includegraphics[width=0.4\textwidth]{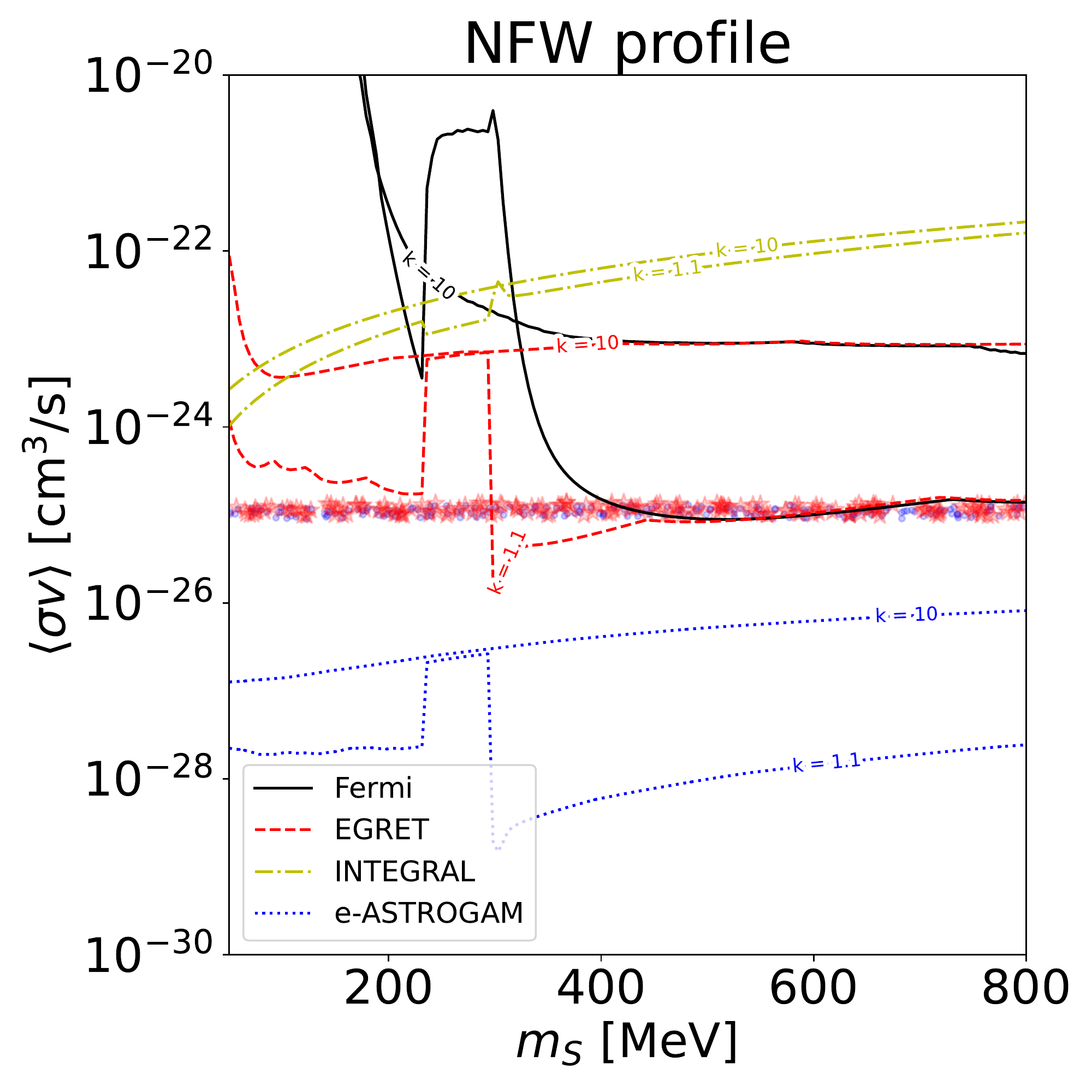}\\
\includegraphics[width=0.4\textwidth]{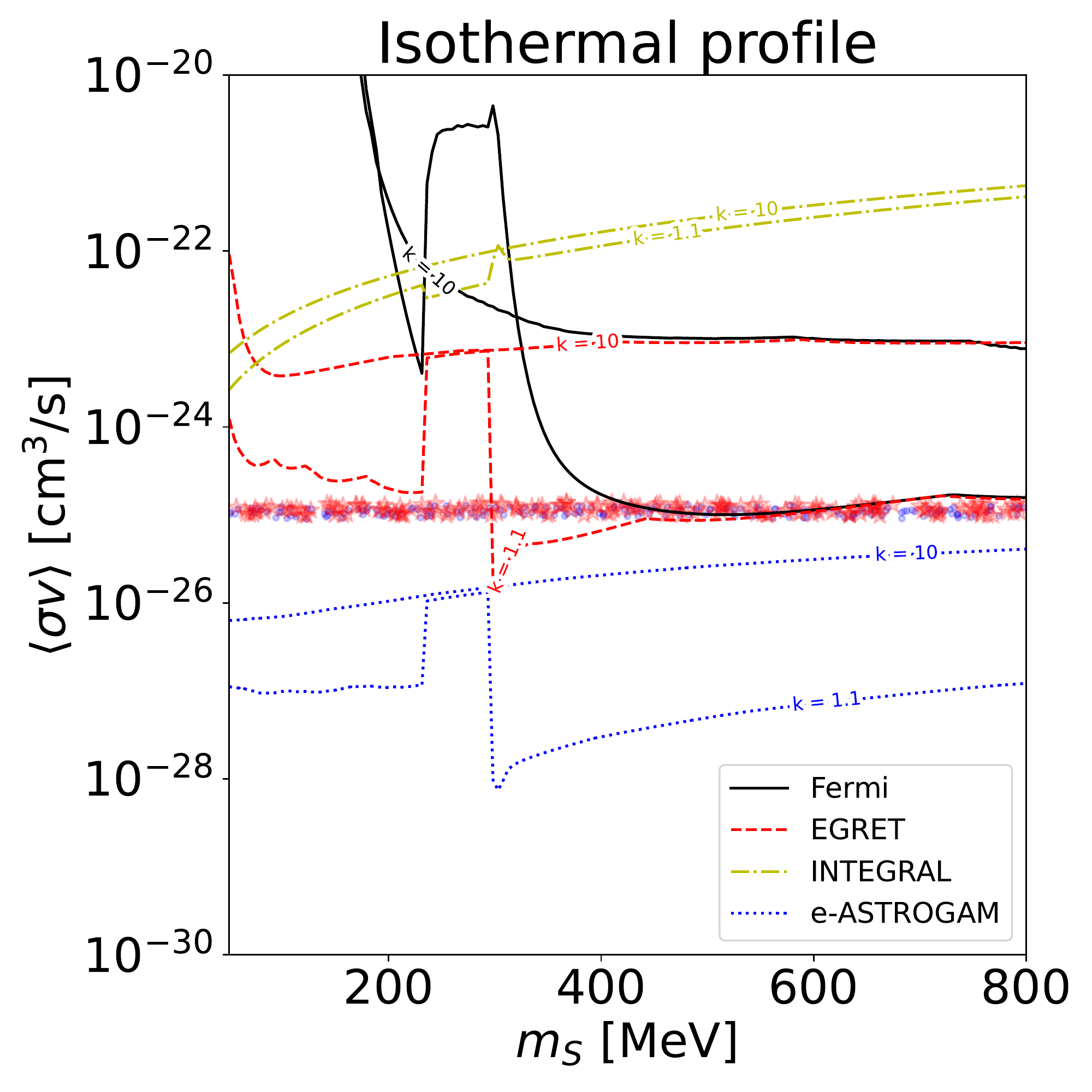}
\caption{We project points in Fig.\ref{fig:asvphi} on two panels, the meaning of the points are the same as Fig.\ref{fig:asvphi}. We choose four representative experiments to set limits on our semi-annihilation DM. INTEGRAL~\cite{WMAP:2012nax} gives almost no limitation to our model, for the low sensitivity of 
COMPTEL~\cite{1998PhDT3K}, we ignore the result.
For the high sensitivity of e-ASTROGAM, all the phenomenology points with correct relic density will be excluded. {We can see that the DM profile difference has a minor impact on the results.}}
\label{fig:excluding_line}
\end{figure}

We study the exclusion limits of two cases: $k=1.1$ and $k=10$ for both the NFW profile and the Isothermal profile. The result is shown in Fig.~\ref{fig:excluding_line}, to learn the limitations clearly, we project points in Fig.~\ref{fig:asvphi} on two panels. {The corresponding limitations on thermal cross-section change mildly and even become weaker compared with the NFW profile.} For the limitation from Fermi-LAT, the constraint is weak in the low mass region $m_S<200$ MeV for both
two cases. This is because Fermi-LAT energy detecting range is 150 MeV - 95 GeV, 
and mass of mediator is $m_\phi<m_S<200$ MeV, meaning the energy of gamma-ray produced by $\phi$ decay is almost out of  Fermi-LAT energy detecting range. At the same time, in mass region $m_S>200$ MeV of case $k=10$, the constrain is still quite weak, this is because in this case, the mediator mass is $m_\phi < 100$ MeV, resulting in limited $\phi$ decay channels and a blunt $\phi$ decay spectrum. But in the case of $k=1.1$ in the mass region $m_S>200$ MeV, the result is quite different, a bulge present on the excluding line around mass region $m_S\in [200, 300]$ MeV, meaning the constrained limit is relaxed, the reason why this happened is that the decay channel $\phi \rightarrow \mu^+ \mu^-$ is opened and it dominant the decay branching fractions, this will suppress high energy final state radiation coming from $\phi\rightarrow e^+e^-$ decay channel, as explained in Fig.~\ref{fig:m_phi1}. When $m_S>300$ MeV ($m_\phi$ is also in this mass region for the highly degenerated mass), the newly opened $\phi\rightarrow p^{0(\pm)} \pi{0(\mp)}$ will dominant and produce a massive number of high energy photon, making the constrain strict again. For the result given by EGRET, the excluding result is similar to that of 
Fermi-LAT, except in low mass region $m_S<200$ MeV, is much more intensive, mainly because EGERT concentrates on the energy range 27 MeV - 8.6 GeV. From Fig.~\ref{fig:excluding_line}, we can see the phenomenology points survived from EGRET and Fermi-LAT exclusion in the case of light $m_\phi$ ($k=10$) of all $m_S$ region, and for the case $k=1.1$, when $m_S>300$ MeV, all points are excluded, while in low $m_S$ region, EGRET excludes parts of region. Unfortunately, all points of the two cases will be excluded by the e-ASTROGAM {future reach}, mainly because e-ASTROGAM is highly sensitive to low-energy gamma-ray signals.

\section{Conclusion}
The traditional WIMP is growing more unrealistic due to DM direct detection's increasingly strict constraints. We consider light DM with $Z_3$ symmetry, which is easy to evade the DM direct detection constraint. Additionally, our $Z_3$  DM annihilates in a different manner: semi-annihilation. In this work, our $Z_3$ DM model contains one complex scalar with $Z_3$ symmetry as a DM candidate and one extra
scalar mediator $\phi$ which mixed with SM Higgs after the SSB of origin $\Phi$, and the Yukawa couplings of  $\phi$-SM are quite SM like. The scalar mediator act as a connecting bridge between DM and SM sector. We only consider DM in the low mass region with $m_S\simeq {\cal O}(100)$ MeV. To learn the specificity of semi-annihilation, we consider the pure semi-annihilation situation by turning off the Higgs portal part and  suppressing the channel $S S\rightarrow \phi \phi$. We consider two cases of $k= m_S/m_{\phi}$ mass ratio, $k=1.1$ highly degenerate case, and $k=10$ extremely light mediator case. Both cases we get the correct relic density in the narrow band $0.094\leq\Omega h^2\leq 0.129$.

In terms of Hamza, we also discuss the MeV gamma-ray produced by present now DM annihilation in the center of the galaxy, which regards an indirect detection signal. The MeV gamma-ray signal comes from the decay of annihilating produced $\phi$, which is highly $m_\phi$ dependent. This is mainly because the available decay channel is strictly related to $m_\phi$, when $m_\phi< 2m_\mu\simeq 100$ MeV, the dominant source of $\phi$ decay spectrum comes from $e^\pm$ final state radiation, but when $m_\phi> 2m_\mu$, for the large decay branching ratio of $\phi\rightarrow \mu^+\mu^-$, the number of hard photon coming from double electron final state radiation will be suppressed, which relax indirect constrain. Since our model is a semi-annihilation one, compare with the traditional model such as the Higgs/$\phi$-portal case, in which DM annihilates into a pair of mediator $\phi$, DM in our model will only generate one mediator $\phi$ with another dark matter $S^*$, suppressing the indirect detection signal. What's more, for the mass difference between annihilation products, the $\phi$ will be less boosted, making another difference compared with the traditional Higgs-portal DM model. We also get the exclusion limits from the existing Fermi-LAT and EGRET, all phenomenology points with correct relic density in $k=10$ case will survive, but in the case of highly degenerate case $k=1.1$, our model receives stringent limitation in the region $m_S>300$ MeV. For the high sensitivity of upcoming e-ASTROGAM, our model would be excluded if there was no observed signal from e-ASTROGAM.

\begin{acknowledgments}
 This work was supported by the National Natural Science Foundation of China under grants No. 12275134, 12275232, and 12005180, by the Natural Science Foundation of Shandong Province under Grant No. ZR2020QA083, and by the Project of Shandong Province Higher Educational Science and Technology Program under Grants No. 2019KJJ007.

\end{acknowledgments}


\end{document}